\documentclass[conference, a4paper]{IEEEtran}
\IEEEoverridecommandlockouts
\usepackage{cite}
\usepackage{amsmath,amssymb,amsfonts}
\usepackage{graphicx}
\usepackage{textcomp}
\usepackage{xcolor}
\usepackage{csquotes}
\usepackage{subcaption}
\usepackage{booktabs}
\usepackage{multirow}

\usepackage{algorithm}
\usepackage{algpseudocode}

\algrenewcommand\algorithmicensure{\textbf{Steps:}}
\algnewcommand{\LineComment}[1]{\State \(\triangleright\) \textit{#1} \hfill~}

\usepackage{url}
\usepackage{siunitx}
\DeclareSIUnit{\belmilliwatt}{Bm}
\DeclareSIUnit{\dBm}{\deci\belmilliwatt}
\DeclareSIUnit{\dBi}{\deci\bel i}
\usepackage[nolist]{acronym} 
\usepackage{xcolor,colortbl}

\usepackage{pifont}%
\usepackage{wasysym}
\newcommand{\cmark}{\CIRCLE}%
\newcommand{\xmark}{\Circle}%
\newcommand{\pmark}{\LEFTCIRCLE}%

\definecolor{DDGray}{gray}{0.8}
\definecolor{DGray}{gray}{0.82}
\definecolor{Gray}{gray}{0.85}
\definecolor{LGray}{gray}{0.9}
\definecolor{LLGray}{gray}{0.95}

\newcolumntype{e}{>{\columncolor{DDGray}}c}
\newcolumntype{f}{>{\columncolor{DGray}}c}
\newcolumntype{g}{>{\columncolor{Gray}}c}
\newcolumntype{h}{>{\columncolor{LGray}}c}
\newcolumntype{i}{>{\columncolor{LLGray}}c}
\usepackage[inline,shortlabels]{enumitem}

\begin{document}
\bstctlcite{IEEEexample:BSTcontrol} 
\title{open5Gcube: A Modular and Usable Framework for Mobile Network Laboratories}

\author{
    \IEEEauthorblockN{
        Thorsten Horstmann\IEEEauthorrefmark{1}\IEEEauthorrefmark{4}, 
	Dominik Brunke\IEEEauthorrefmark{3},
        Tobias Kremeyer\IEEEauthorrefmark{1}\IEEEauthorrefmark{4},
        Matthias Wilmes\IEEEauthorrefmark{1}, 
        Gunnar Schneider\IEEEauthorrefmark{1}, \\
        Julian Sturm\IEEEauthorrefmark{3},
        Hartmut König\IEEEauthorrefmark{3},
 	Michael Rademacher\IEEEauthorrefmark{1}\IEEEauthorrefmark{4}  
    }
    \IEEEauthorblockA{
        \IEEEauthorrefmark{1}Fraunhofer FKIE, Bonn, Germany, {firstname.lastname}@fkie.fraunhofer.de}
	\IEEEauthorblockA{\IEEEauthorrefmark{3}ZITiS, Munich, Germany {firstname.lastname}@zitis.bund.de}
        \IEEEauthorblockA{\IEEEauthorrefmark{4}Hochschule Bonn Rhein-Sieg, Sankt Augustin, Germany, {firstname.lastname}@h-brs.de}
}

\maketitle

\begin{abstract}
In mobile network research, the integration of real-world components such as User Equipment (UE) with open-source network infrastructure is essential yet challenging. To address these issues, we introduce open5Gcube, a modular framework designed to integrate popular open-source mobile network projects into a unified management environment. Our publicly available framework allows researchers to flexibly combine different open-source implementations, including different versions, and simplifies experimental setups through containerization and lightweight orchestration. We demonstrate the practical usability of open5Gcube by evaluating its compatibility with various commercial off-the-shelf (COTS) smartphones and modems across multiple mobile generations (2G, 4G, and 5G). The results underline the versatility and reproducibility of our approach, significantly advancing the accessibility of rigorous experimentation in mobile network laboratories.
\end{abstract}

\begin{IEEEkeywords}
5G, Testbed, Open-Source
\end{IEEEkeywords}

\section{Introduction and Motivation}\label{sec:motivation}
In mobile network research, a challenge is the integration of various \ac{RAN} and \ac{CN} components for usage with real-world \ac{UE}. This integration is needed to demonstrate ideas for future network generations or to uncover security vulnerabilities in existing standards. There are two worlds of mobile network components: Commercial solutions and open-source alternatives. Both come with advantages and disadvantages. 

Commercial, closed-source systems are often expensive, and they do not provide the interfaces or modification capabilities required for in-depth experimental analysis. However, they (usually) offer a high stability and good compatibility. In contrast, open-source solutions offer flexibility and transparency that facilitate customization. However, these systems come with challenges, including a steep learning curve and potential compatibility issues with \ac{COTS} \acp{UE}. This dichotomy highlights the need for a \textit{modular and usable framework for mobile network laboratories} that combines the adaptability of \ac{OSS} with the ease-of-use necessary for experimental setups.

From a security perspective, commercial systems frequently operate as black boxes, constraining access to internal mechanisms and impeding detailed vulnerability assessments. In contrast, open-source projects provide unrestricted access to source code, enabling researchers to scrutinize and modify implementations. This flexibility permits the development of robust proof-of-concept models for simulating attack vectors and assessing detection methods, thereby facilitating systematic evaluations and improvements of security measures within mobile communication frameworks. Such transparency is integral to advancing security protocols and ensuring resilience against emergent threat paradigms in 5G networks.

In this work, we introduce open5Gcube, a framework to enable researchers to conduct rigorous and reproducible experiments in mobile network research. Compared to previous work (see Section~\ref{sec:previous_work}), our framework 
\begin{enumerate*}[1)]
\item integrates the most common open-source mobile network projects into a unified management framework,
\item allows researchers to use these projects in different versions, and 
\item was successfully used to test the compatibility of numerous \ac{COTS} \acp{UE}.
\end{enumerate*}

\section{Previous Work}\label{sec:previous_work}

Since testbeds play a crucial role in mobile network research, several publications have reported on their design and usage, each focusing on different aspects~\cite{Schauer, amini_5g_2023, amini_comparative_2024, bozis_versatile_2024, bozis_enhancing_2024, chepkoech_evaluation_2023, hakegard_performance_2024, john_reference_2022, reddy_open_2023, mamushiane_deploying_2023, sahbafard_performance_2023, seidel_how_2023, valente_easy_2024, vanichchanunt_implementation_2024}. Multiple contributions provide detailed practical descriptions of the setup and configuration of a 5G \ac{SA} network using \ac{OSS}, highlighting central configuration parameters~\cite{john_reference_2022, mamushiane_deploying_2023, seidel_how_2023}. Other studies set up a 5G \ac{SA} deployment and, using this deployment, perform measurements to evaluate the effect of different configuration parameters on relevant performance metrics, such as latency, throughput, and received signal power~\cite{sahbafard_performance_2023, hakegard_performance_2024, bozis_versatile_2024, bozis_enhancing_2024}. A further significant body of research concentrates on comparative analyses of different 5G and 4G deployments, measuring the effect of different combinations of open-source \acp{BS} and \acp{CN} on performance~\cite{amini_5g_2023,  amini_comparative_2024, chepkoech_evaluation_2023, reddy_open_2023}. Additionally, some studies implement advanced use cases for 5G networks using \ac{OSS} projects, such as \acp{RAN} based on Unmanned Aerial Vehicles~\cite{valente_easy_2024} and network slicing aimed at very low latency services~\cite{vanichchanunt_implementation_2024}.

\begin{table*}[]
\caption{Overview of related work with a focus on used software and hardware.}
\resizebox{\textwidth}{!}{%
\setlength{\tabcolsep}{2pt}
\renewcommand{\arraystretch}{0.8}
\begin{tabular}{c|c|c|c|ccc|cccccc|cccc}
\label{tab:overview}
\multirow{3}{*}{Paper} & \multirow{3}{*}{Year} & \multirow{3}{*}{\begin{tabular}[c]{@{}c@{}}Unified \\ Manag.\end{tabular}} & \multirow{3}{*}{\begin{tabular}[c]{@{}c@{}}Config.\\ available\end{tabular}} & \multicolumn{3}{c|}{\multirow{2}{*}{UE}} & \multicolumn{6}{c|}{RAN} & \multicolumn{4}{c}{\multirow{2}{*}{Core}} \\
 &  &  &  & \multicolumn{3}{c|}{} & \multicolumn{4}{c|}{Software} & \multicolumn{2}{c|}{Hardware} &  \\
 &  &  &  & Phones & Modem & Soft & srs & OAI & UERS. & \multicolumn{1}{c|}{Osmo} & SDR & COTS & srs & Open5GS & Free5GC & OAI \\
 \midrule
\cite{amini_5g_2023} & 2023 &\xmark&\xmark&\xmark&\cmark &\xmark&\cmark &\xmark&\xmark& \multicolumn{1}{c|}{\xmark} &\cmark &\xmark&\xmark&\cmark &\xmark&\xmark\\
\rowcolor{LLGray} 
\cite{amini_comparative_2024} & 2024 &\xmark&\xmark&\xmark&\cmark &\xmark&\cmark &\cmark &\xmark& \multicolumn{1}{c|}{\xmark} &\cmark &\xmark&\xmark&\cmark &\xmark&\cmark \\
\cite{bozis_versatile_2024},\cite{bozis_enhancing_2024} & 2024 & \pmark & \cmark & \xmark & \xmark & \cmark & \xmark & \cmark & \xmark & \multicolumn{1}{c|}{\xmark} & \cmark & \xmark & \xmark & \xmark & \xmark & \cmark \\
\rowcolor{LLGray} 
\cite{chepkoech_evaluation_2023} & 2023 &\xmark&\xmark&\cmark &\cmark &\cmark &\cmark &\cmark &\xmark& \multicolumn{1}{c|}{\xmark} &\cmark &\cmark &\cmark &\cmark &\xmark&\cmark \\
\cite{hakegard_performance_2024} & 2024 &\xmark&\xmark&\cmark &\cmark &\cmark &\cmark &\xmark&\xmark& \multicolumn{1}{c|}{\xmark} &\cmark &\xmark&\xmark&\cmark &\xmark&\xmark\\
\rowcolor{LLGray} 
\cite{john_reference_2022} & 2022 &\pmark&\cmark &\xmark&\cmark &\xmark&\xmark&\cmark &\xmark& \multicolumn{1}{c|}{\xmark} &\cmark &\xmark&\xmark&\xmark&\xmark&\cmark \\
\cite{reddy_open_2023} & 2023 &\pmark &\xmark&\xmark&\xmark&\cmark &\xmark&\xmark&\cmark & \multicolumn{1}{c|}{\xmark} &\xmark&\xmark&\xmark&\cmark &\cmark &\cmark \\
\rowcolor{LLGray} 
\cite{mamushiane_deploying_2023} & 2024 &\pmark&\xmark&\cmark &\cmark &\cmark &\cmark &\xmark&\xmark& \multicolumn{1}{c|}{\xmark} &\cmark &\xmark&\xmark&\cmark &\xmark&\xmark\\
\cite{sahbafard_performance_2023} & 2023 &\xmark&\xmark&\xmark&\cmark &\xmark&\xmark&\cmark &\xmark& \multicolumn{1}{c|}{\xmark} &\cmark &\xmark&\xmark&\xmark&\xmark&\cmark \\
\rowcolor{LLGray} 
\cite{seidel_how_2023} & 2023 &\xmark&\xmark&\xmark&\cmark &\cmark &\xmark&\cmark &\xmark& \multicolumn{1}{c|}{\xmark} &\cmark &\xmark&\xmark&\xmark&\xmark&\cmark \\
\cite{valente_easy_2024} & 2024 &\xmark&\xmark&\cmark &\cmark &\cmark &\cmark &\xmark&\xmark& \multicolumn{1}{c|}{\xmark} &\cmark &\xmark&\xmark&\cmark &\xmark&\xmark\\
\rowcolor{LLGray} 
\cite{vanichchanunt_implementation_2024} & 2024 &\xmark&\xmark&\cmark &\xmark&\xmark&\cmark &\xmark&\xmark& \multicolumn{1}{c|}{\xmark} &\cmark &\xmark&\xmark&\cmark &\xmark&\xmark\\
This & 2025 &\cmark &\cmark &\cmark &\cmark &\cmark &\cmark &\cmark &\cmark & \multicolumn{1}{c|}{\cmark} &\cmark &\cmark &\cmark &\cmark &\cmark &\cmark
\end{tabular}
}
\end{table*}

Table~\ref{tab:overview} provides a concise summary, highlighting software and hardware utilized across different publications. Please note that this list does not constitute an exhaustive survey. We present the ones we consider the most relevant and recent.

A mobile network laboratory, like real-world deployments, consists of three components: \acp{UE}, \acp{BS}, and the \ac{CN}.

For \textbf{\aclp{UE}}, previous studies identified three types: \ac{COTS} smartphones; modems equipped with embedded boards; and so-called Soft-\acp{UE}, which are purely software-based implementations running either on \acp{SDR} or in emulated environments. As visualized in Table~\ref{tab:overview}, most previous work utilizes programmable modems due to their reported compatibility with various \ac{RAN} and \ac{CN} implementations. Smartphones and Soft-\acp{UE} are also commonly used. However, it has been noted that smartphone compatibility is not universal across open-source projects. It remains unclear whether incompatibilities primarily stem from the operating system (e.g., Android or iOS), baseband chipset (e.g., Exynos, Qualcomm, or MediaTek), or specific configuration parameters (e.g., frequency bands)~\cite{srsran_cots_ues:online}. Soft-\acp{UE} are useful for evaluating novel concepts such as new \ac{RAN}-protocols; however, they lack real-world comparability~\cite{Rouili}.

Implementing a \textbf{\acl{BS}} requires both hardware and software, typically \ac{OSS}. Regarding hardware choices for base stations (\acp{BS}), most of previous work relies on \acp{SDR}, as indicated in Table~\ref{tab:overview}. Only a few projects incorporate commercial base stations (\ac{COTS} \acp{BS}), likely due to higher costs and limited adaptability. Four \ac{OSS} projects are predominantly utilized: srsRAN (srsRAN Project and srsRAN 4G)~\cite{srsRAN-Online}, \ac{OAI}~\cite{OAI-Online}, UERANSIM~\cite{UERANSIM-Online}, and OSMO-BTS~\cite{OSMO-Online}. Among these projects, srsRAN and \ac{OAI} are most frequently employed within our evaluated subset of literature. OSMO-BTS is less commonly considered since it implements a 2G network; however, given well-known vulnerabilities and ongoing global availability of 2G networks, we include it~\cite{cattaneo2013review}.

A similar landscape of \ac{OSS} exists for the \textbf{Core Network}. Implementations can be directly related to the \ac{BS} software - such as srsRAN~\cite{srsRAN-Online} or \ac{OAI}~\cite{OAI-Online} - or independent projects like Free5GC~\cite{free5gc-Online} or Open5GS~\cite{Open5GS-Online}, which focus on the \ac{CN}. 

In the context of this work, we emphasize two major aspects for \textbf{usability} and \textbf{reproducibility} of experimental testbeds:

\begin{itemize}[leftmargin=*]
	\item \textbf{Unified Management:} The presence of a framework facilitating orchestration across multiple projects to simplify the setup and maintenance (usability).
	\item \textbf{Configuration Availability:} Accessibility of project configurations provided as supplementary material alongside publications (reproducibility).
\end{itemize}

Surprisingly, both aspects remain largely unaddressed in the existing literature (see Table~\ref{tab:overview}). Only a few publications have made their configurations publicly accessible~\cite{bozis_versatile_2024,john_reference_2022}. Even less common is the implementation of unified management frameworks. Publications partially addressing this issue are discussed in the following. In~\cite{john_reference_2022}, authors describe using Docker containers for automated setup and provide a detailed tutorial on configuring various components; however, their referenced repository is not publicly accessible. Mamushiane et al.~\cite{mamushiane_deploying_2023} describe utilizing OpenStack with multiple \ac{CN} implementations but provide limited details regarding their software infrastructure. The study by Reddy et al.~\cite{reddy_open_2023} employs VirtualBox's graphical user interface as a basic management tool. Finally, Bozis et al.~\cite{bozis_versatile_2024, bozis_enhancing_2024} present an evolving laboratory setup at different development stages along with scripts to configure an \ac{OAI}-based laboratory environment.

As summarized at the bottom of Table~\ref{tab:overview} and elaborated in subsequent sections, we contribute a framework implementing a unified management which we provide open-source alongside our configurations.

\section{Architecture}
\label{sec:architecure}
\begin{figure*}[h]
\centering
\includegraphics[width=\linewidth]{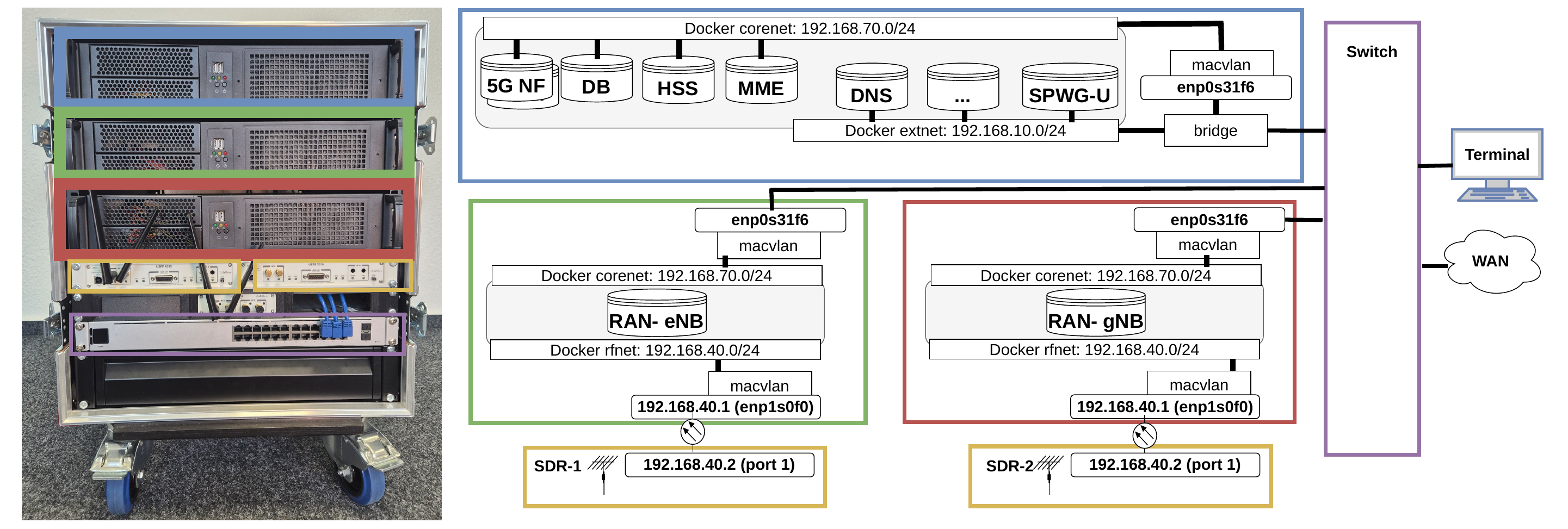}
\caption{Representation of our open5Gcube. On the left side, the components in the 19" rack. On the right side, the corresponding systems and their interconnections are depicted.}
\label{fig:design_cube}
\end{figure*}

The purpose of this section is to describe the design choices underlying our framework. We divided this section into two main parts: Hardware Design and Software Design.

\subsection{Hardware Design}
The primary objective of our hardware design is \textit{flexibility}. Therefore, we divided the system into individual, interchangeable hardware components. Additionally, \textit{mobility} is essential, as it allows the system to be easily transported to testing locations by fitting into a standard car. Another critical hardware requirement are \textit{dedicated computing resources} in the radio domain to address high single-thread loads observed in individual processes. Thus, each computer requires a powerful CPU, which consequently generates significant heat. To ensure broad software compatibility, we selected the widely used \ac{USRP} X310 and \ac{USRP} B210 \acp{SDR} from Ettus Research. The X310 \ac{SDR} requires a 10G connection. Furthermore, precise synchronization among \acp{BS} in mobile networks is essential for communication between components; therefore, an external clock source is required. We transferred those requirements into a prototype which is shown in Fig.~\ref{fig:design_cube}, and will be further detailed below.

\textbf{Reference Hardware Setup:} 
Our design includes three subunits: one \ac{CN} and two \acp{BS}, supporting numerous typical scenarios, including a 5G \ac{NSA} scenario with one \ac{eNB} and one \ac{gNB}. Each subunit uses an identical computer system for self-containment and interchangeability. To facilitate mobility and be able to transport our system in a car, we use a mobile server rack in a 19-inch flight-case (max 12 \ac{U}) with compact 2\ac{U} servers featuring Intel i9 11900K CPUs and 10G network cards. A preconfigured 1\ac{U} switch handles network connections.

The rack includes two X310 \acp{SDR} with \ac{UBX} daughterboards connected via 10G cables, a Quectel RM500Q-GL Development Kit, and optional space for B210 \acp{SDR}. A Leo Bodnar \ac{GPSDO} provides a stable 10 MHz reference signal for the \acp{SDR}, maintaining output even without a GPS fix \cite{leo_bodnar}.

\textbf{Core System and Kernel:} For maximal performance, hyper-threading, CPU frequency control, C-States, P-States and any other power management settings were disabled from the BIOS. Ubuntu 22.04 was selected as the operating system due to its compatibility with software package requirements of individual \acp{CN}. We installed and enabled a low-latency kernel (5.15.0-133) ensuring reliably low system response times essential for real-time performance in 5G applications. We observed that industrial products such as Amarisoft \cite{amarisoft} employ similar operating systems and kernel configurations.

\textbf{Network:} The network topology consists of three interconnected systems communicating via Ethernet (1G) through the switch. Additionally, two \ac{RAN} hosts establish direct 10G connections with one X310 \ac{SDR} each.

As described in the following subsection~\ref{sec:architecture_software}, services are deployed as Docker containers. Thus, the network architecture incorporates three virtual Docker networks:
\begin{itemize}[leftmargin=*]
    \item \texttt{corenet}: Configured as a macvlan~\cite{docker_macvlan} network, this network is designated for all the \ac{CN} components. The use of 802.1q trunk bridge mode allows flexible distribution of containers across multiple servers. The host system is configured with additional virtual interfaces, facilitating direct communication with services within the network.
    \item \texttt{extnet}: Bridged to the WAN, this network facilitates the \ac{CN} Gateway (SPGW-U, UPF) and other services (like DNS) for external connectivity through the host interface.
    \item \texttt{rfnet}: This network connects the \ac{USRP} X310 \acp{SDR}, providing an isolated network for their operations.
\end{itemize}

\subsection{Software and Management}\label{sec:architecture_software}
\begin{figure}[h!]
	\centering
	\vspace{-3ex}
	\includegraphics[width=1\linewidth]{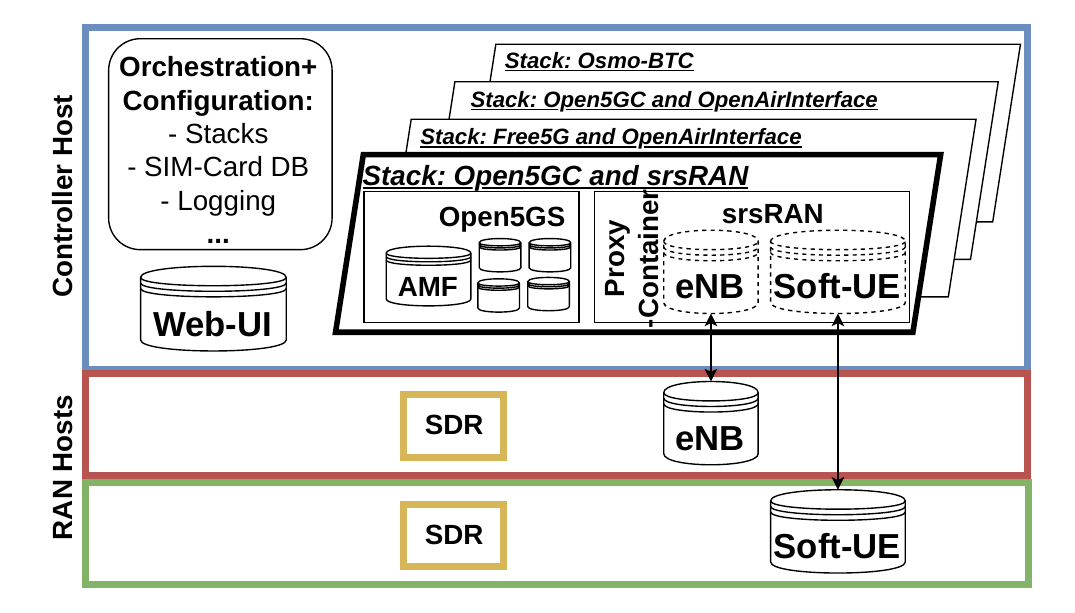}
	\caption{Logical software organization of the framework. Open-source projects (e.g., Open5GS, srsRAN) are deployed by our framework on different host and summarized into \textit{stacks}.}
\label{fig:cube-software-orga}
\end{figure}

In this subsection, we describe the main design choices and functionality of our software framework, which runs on the reference setup described in the previous subsection. A visualization of the software framework is shown in Figure~\ref{fig:cube-software-orga}.

\textbf{Containerization:} Our framework utilizes container virtualization based on Docker. Using containers, services (e.g., eNB, gNB, 5G \acp{NF}) run in isolated environments, providing benefits such as streamlined deployment, reduced dependency conflicts, and consistent installations across setups. Multiple services are grouped and organized into use-case-specific \textbf{stacks} (e.g., Open5GC-srsRAN), conveniently defined and managed using Docker-Compose files. Docker-Compose provides a declarative and straightforward mechanism to specify container configurations and interdependencies (\ac{RAN}, \ac{CN} \acp{NF}, databases, etc.), simplifying the setup and reproducibility of complex network scenarios involving different open-source projects.

Although the images are not optimized for size, they closely follow the structure of the official images associated with their respective projects as far as reasonable. All images are built upon a common base image. This approach ensures consistency and incorporates essential configurations. Additionally, the base image provides a user-friendly environment within a running container, facilitating convenient console interactions. It includes an up-to-date Wireshark version to easily trace and analyze network traffic inside the containers.

\textbf{Orchestration:} For orchestrating services across multiple hosts, two solutions were thoroughly evaluated: Kubernetes and Docker Swarm. Kubernetes, a widely recognized industry standard for container orchestration, inherently supports cloud-native setups and is adopted by \acp{MNO} for 5G deployments due to its scalable architecture. Despite these advantages, Kubernetes exhibits substantial drawbacks for laboratory applications, including a steep learning curve and complex management largely unnecessary in smaller-scale laboratory settings. \cite{kubernetes} In contrast, Docker Swarm—Docker's native orchestration solution—integrates seamlessly with Docker-Compose, providing straightforward development for laboratory-scale environments. 
\cite{docker_swarm}

However, a drawback shared by both Kubernetes and Docker Swarm is their conceptual overhead. From our experience, we found that neither system adequately represents our laboratory scenario. For instance, both solutions provide powerful scaling and load balancing mechanisms, which we do not require. On the other hand, with both solutions, assigning static fixed IP addresses to services is complex and needs extensive workarounds. This is because containers are considered as ephemeral and the networking is dynamically managed by these orchestration tools. Nevertheless, some of the open-source projects integrated in our framework require IP addresses rather than hostnames in their configuration files (e.g. OAI-RAN, Kamailio). Determining the IP addresses in such network setups at runtime would increase the integration effort significantly.
We also consider a setup with fixed IP addresses easier to use and debug and especially more reproducible. In consequence, we argue that neither of both solutions is a good fit to our requirements.

Considering these trade-offs, we developed a \textbf{simplified orchestration mechanism}, tailored specifically for small-scale mobile network laboratory setups. Our approach employs a script-based implementation to minimize complexity. One host—termed the \enquote{Controller Host} (cf.\ Figure~\ref{fig:cube-software-orga}) centrally manages and initiates all containerized services. This host contains the project repository and operates all stack services (particularly those related to the \ac{CN}), while delegating the execution of \ac{RAN} containers to configured \enquote{RAN Hosts}.

In more detail, the orchestration workflow functions as follows: Upon starting a stack on the Controller Host, RAN-related containers (such as \ac{eNB} or \ac{gNB}) are also initiated. The entry-point script of these containers includes logic to detect whether the current host matches the configured target host. Using \ac{DooD}, this script (executed by container entry-point) performs the following operations:

\begin{algorithm}[h!]
\footnotesize
\label{alg}
\begin{algorithmic}
\Require
\State \textbullet~Docker client is installed in the image 
\State \textbullet~Docker host socket is mounted into the container
\Ensure
\State \textbullet~Determine host machine name
\If{container is running on the configured RAN host}
\State Start RAN component and \Return
\EndIf
\LineComment{Release all assigned static IP addresses}
\ForAll{connected network interfaces}
\State disconnect network from container
\EndFor
\State \textbullet~Connect the container to the default Docker bridge
\State \textbullet~Transfer necessary files to the remote host via SSH
\State \textbullet~Execute Docker-Compose on the remote daemon via SSH %
\end{algorithmic}
\end{algorithm}
\vspace{-1ex}

This lightweight strategy leverages standard Docker-Compose features, extended by minimal scripting, and effectively simplifies remote deployment. 

\textbf{Networking:} To effectively manage networking, Docker's built-in networking capabilities are utilized. The \texttt{corenet}, which is configured to use macvlan networking, allows containers to be attached directly to physical network interfaces with unique MAC addresses. This approach enables direct communication between services over the network without requiring additional routing or bridging setups. Consequently, services can easily be distributed across multiple physical hosts, significantly reducing complexity typically associated with multi-host network management.

\textbf{Configuration:} To facilitate the configuration of the stacks, we use simple central environment files which can be processed natively by Docker. Global settings such as channel number, frequency band, cell-Id etc. as well as the fixed IP addresses of the services are set centrally and used by the Docker-Compose files. 
We provide configuration files for all integrated projects where these parameters are defined as template variables. 
When the services are started, these settings are substituted in the configuration templates of the projects. 
Stacks can reuse the provided project configuration files and global settings or overwrite them if needed.
The \ac{IMSI} and corresponding SIM card keys are specified in an environment file, too. On startup of a stack the particular subscriber database is populated from scratch to always have a clean state for the experiments.

\textbf{Management:} In addition to command-line-based interactions, we developed a simple Web-UI for interacting with the framework. Its principal use cases include configuring, starting and stopping specific stacks, and a live display of running containers and their logs, using color-coding to indicate the health status of the currently running stack.

The primary objective of our framework is to facilitate flexible combinations and rapid deployment of various open-source \ac{CN}, \ac{RAN}, and \ac{UE} implementations. By providing a modular repository that includes prominent and commonly used open-source projects within the mobile communications community, different implementation combinations can be easily tested, as demonstrated in the following section.

\section{Evaluation}
\label{sec:evaluation}
In this section, we present the results of experiments that can be conducted using the open5Gcube framework. Although our framework is application-driven and designed to realize specific mobile network scenarios (e.g., testing particular vulnerabilities, evaluating roaming scenarios, etc.), this section focuses primarily on results related to \ac{UE} compatibility. To demonstrate the versatile nature of our framework, we evaluated all relevant mobile generations (2G, 4G, and 5G) using a fixed set of different UEs, as detailed in Table~\ref{tab:overview_of_ues_of_evalulation}.

We chose five different \ac{COTS} smartphones from different vendors and a modem that has been commonly used in previous work (see Section~\ref{sec:previous_work}). 

\begin{table}[h]
\caption{Overview of used UEs for testing the stacks.}
\label{tab:overview_of_ues_of_evalulation}
\setlength{\tabcolsep}{2pt}
\centering
\begin{tabular}{c|c|c|c}
                  Index  & Model                & OS Version  & Modem           \\ \midrule
                  S25    & Samsung Galaxy S25   & Android 15  & Snapdragon X80  \\
\rowcolor{LLGray} OP13   & OnePlus 13           & Android 15  & Snapdragon X80  \\
                  ME50P   & Motorola Edge 50 Pro & Android 14  & Snapdragon X63  \\
\rowcolor{LLGray} ARP6   & Asus ROG Phone 6     & Android 14  & Snapdragon X65  \\
                  GP9    & Google Pixel 9       & Android 15  & Exynos 5400     \\
\rowcolor{LLGray} AIP13  & Apple iPhone 13 Mini & iOS 18.4    & Snapdragon X60  \\
                  QRM500 & Quectel RM500Q-GL    & ABR11A06M4G & Snapdragon X55  \\
\end{tabular}
\end{table}

\subsection{2G}
To build-up a 2G network, we defined a stack using the Osmocom implementation for cellular network infrastructure \cite{OSMO-Online}. This stack includes the necessary subprojects to provide \ac{GPRS} for IP connections. Each component is encapsulated in a separated Docker container. The \ac{BTS} uses OsmoTRX connected to the \ac{USRP} SDR as a back-end.
We were able to connect all smartphones shown in Table~\ref{tab:overview_of_ues_of_evalulation}. Telephony, SMS, and the data packet service operated without any issues.

\subsection{4G + VoLTE}
We evaluated our framework in a complex 4G scenario, which includes VoLTE and SMS services. The stack is based on the configuration described in \cite{docker_open5gs} and illustrates the advantages of our
modular approach, where the required projects can be flexibly combined. While the eNB is implemented by srsRAN, the \ac{CN} with \ac{HSS} and \ac{MME} is integrated with Open5GS. The \ac{IMS} is provided by the Kamailio SIP server project \cite{kamailio} and rtpengine \cite{rtpengine}. The SMS service can be handled directly in the \ac{IMS} or optionally provided via SMS-over-SGs by the \ac{MSC} of a parallel running 2G \ac{CN}.

All tested \acp{UE} of Table~\ref{tab:overview_of_ues_of_evalulation} connected to the network and were able to establish a stable IP connection. In contrast, only three \acp{UE} connected successfully to the \ac{IMS}. The problem of establishing a registration is covered by numerous issue entries in \cite{docker_open5gs} and emphasizes the complexity of the VoLTE registration process \cite{volte_mess:online}. Table~\ref{tab:volte_results} summarizes the results. Although VoLTE support is limited for various \acp{UE}, numerous experiments focus solely on a stable 4G data link.

\begin{table}[h!]
\caption{UE Test results for 4G VoLTE stack.}
\label{tab:volte_results}
\setlength{\tabcolsep}{4.2pt}
\begin{tabular}{c|c|c|c|c|c|c|c}
                                   & S25    & OP13   & ME50P  & ARP6   & GP9    & AIP13 & QRM500 \\ \midrule
                  4G data link     & \cmark & \cmark & \cmark & \cmark & \cmark &  \cmark     & \cmark \\
\rowcolor{LLGray} VoLTE & & & & & & &\\
\rowcolor{LLGray} registered & \multirow{-2}{*}{\xmark} & \multirow{-2}{*}{\cmark} & \multirow{-2}{*}{\xmark} & \multirow{-2}{*}{\cmark} & \multirow{-2}{*}{\xmark} &  \multirow{-2}{*}{\cmark}     & \multirow{-2}{*}{—}      \\
\end{tabular}
\end{table}

\subsection{5G SA}
To explore the flexibility of our framework, we configured stacks to build 5G \ac{SA} networks using open-source projects (see Table~\ref{tab:5g_stack_versions}). Each \ac{CN} implementation (Open5GS, OAI-Core, Free5GC) is combined with a \ac{RAN} project (srsRAN, OAI-RAN) and tested separately with all \acp{UE} mentioned in Table~\ref{tab:overview_of_ues_of_evalulation}. In addition, each \ac{CN} was connected to UERANSIM for an emulated setup. 
By using the global settings and configuration template files, as described in section~\ref{sec:architecture_software}, it is not necessary to define any stack-specific configuration. So, a stack definition is simplified to just referencing the necessary services in a Docker-Compose file.

\begin{table}[h!]
\caption{Versions of the projects used in the stacks.}
\label{tab:5g_stack_versions}
\centering
\begin{tabular}{c|c|c|c|c}
srsRAN  & OAI-RAN  & Open5GS & OAI-Core & Free5GC \\ \midrule
24.10.1 & 2024-w23 & 2.7.2   & 2.1.0    & 3.4.4
\end{tabular}
\end{table}

Table~\ref{tab:5g_sa_results} gives a summary of the test results.
The Samsung Galaxy S25, OnePlus 13, Asus ROG Phone 6 and the Quectel RM500Q-GL worked with all stacks. Each \ac{UE} connected to the \ac{RAN} was authenticated by the \ac{CN} and successfully established a \ac{PDU} session. To stress test the link, a flood-ping was performed for two minutes without any connection loss.

The Motorola Edge 50 Pro connects to the \acp{RAN} and establishes \ac{PDU} sessions as well, but disconnects itself from the network after a few seconds due to the missing \ac{IMS}. This is a known problem with some combinations of \ac{UE} model and Android version \cite{srsran_cots_ues:online} which could possibly be prevented by specific Carrier Setting configurations (e.g.\ with the OpenMobileNetworkToolkit \cite{omnt:online}) or by disabling the feature with a hidden IMS settings menu.

In our tests, the Google Pixel 9 was unable to connect to srsRAN, as the cell itself was not detected by the device. With OAI-RAN, a link was established, but in combination with Free5GC no \ac{PDU} session was created. With the two other \acp{CN}, the tests passed without any problems.

For the iPhone, the settings have been adjusted. In particular, the \ac{SUPI} concealment was activated on the Sysmocom SIM card and configured in Open5GS and Free5GC. In combination with srsRAN and OAI-RAN the iPhone connected to the private 5G network. Since the OAI-Core does not support SUCI-based authentication, the iPhone could not connect to it.

\begin{table}[h!]
\caption{UE Test results for 5G SA stacks.}
\label{tab:5g_sa_results}
\setlength{\tabcolsep}{2.4pt}
\begin{tabular}{c|l|c|c|c|c|c|c|c}
\multicolumn{2}{c|}{Stack}            & S25    & OP13   & ME50   & ARP6   & GP9    & AIP13 & QRM500 \\ \midrule
\multirow{3}{*}{srsRAN}    & Open5GS  & \cmark & \cmark & \pmark & \cmark & \xmark & \cmark & \cmark \\
                           & OAI-Core & \cmark & \cmark & \pmark & \cmark & \xmark & \xmark & \cmark \\
                           & Free5GC  & \cmark & \cmark & \pmark & \cmark & \xmark & \cmark & \cmark \\
\rowcolor{LLGray}          & Open5GS  & \cmark & \cmark & \pmark & \cmark & \cmark & \cmark & \cmark \\
\rowcolor{LLGray}          & OAI-Core & \cmark & \cmark & \pmark & \cmark & \cmark & \xmark & \cmark \\
\rowcolor{LLGray}
\multirow{-3}{*}{OAI-RAN}  & Free5GC  & \cmark & \cmark & \pmark & \cmark & \pmark & \cmark & \cmark
\end{tabular}
\end{table}

In conclusion, the tests confirm that---from our perspective---the compatibility of the tested \acp{UE} is adequate for mobile network laboratories. During our experiments, we found that it was particularly helpful to switch quickly between stacks, to always have a consistent and reproducible setup and to simply change the versions of projects used.

\section{Summary}
Realistic testbeds are essential in mobile network research, particularly for security-related work. Commercial solutions lack the necessary flexibility while open-source alternatives tend to be complex to set up and maintain. To address this gap, \textbf{we developed open5Gcube, a modular framework that integrates prominent open-source projects into a unified management environment}. Leveraging Docker containerization, our architecture isolates services to simplify deployment and experimentation. Our prototype hardware setup consists of identical high-performance servers, \acp{SDR} and dedicated 10G networking infrastructure housed in a portable 19-inch rack optimized for transportability among laboratories. Extensive compatibility tests were conducted across multiple mobile generations (2G via Osmocom; 4G VoLTE via srsRAN/Open5GS; 5G SA using combinations of srsRAN, OAI-RAN, Free5GC, Open5GS) using various COTS smartphones and modems. The results demonstrate broad interoperability for basic data connectivity but selective compatibility for advanced features such as VoLTE registration. To encourage reproducibility and collaboration within the research community, \textbf{we have released open5Gcube as an open-source project}~\cite{open5GcubeRepo}, allowing other groups to easily build their experiments and applications upon our framework.

\bibliographystyle{IEEEtran} %
\bstctlcite{IEEEexample:BSTcontrol} %
\bibliography{ref.bib}

\begin{acronym}[]
\acro{3GPP}{3rd Generation Partnership Project}
\acro{SBI}{Service-Based Interface}
\acro{SBA}{Service-Based Architecture}
\acro{NF}{Network Function}
\acro{AMF}{Access \& Mobility Management Function}
\acro{AUSF}{Authentication Server Function}
\acro{COTS}{commercial off-the-shelf}
\acro{NRF}{Network Repository Function}
\acro{SMF}{Session Management Function}
\acro{UDM}{Unified Data Management}
\acro{MNO}{Mobile Network Operator}
\acro{SUPI}{Subscription Permanent Identifier}
\acro{5GC}{5G Core Network}
\acro{OAI}{OpenAirInterface}
\acro{SD}{Slice Differentiator}
\acro{UE}{User Equipment}
\acro{RAN}{Radio Access Network}
\acro{SDN}{Software-Defined Networking}
\acro{NFV}{Network Functions Virtualization}
\acro{MCC}{Mobile Country Code}
\acro{MNC}{Mobile Network Code}
\acro{RAN}{Radio Access Network}
\acro{PDU}{Packet Data Unit}
\acro{UE}{User Equipment}
\acro{SDR}{Software-Defined Radio}
\acro{OSS}{Open source software}
\acro{BS}{Base Station}
\acro{OAI}{OpenAirInterface}
\acro{srs}{TODO}
\acro{SA}{Standalone}
\acro{gNB}{gNodeB}
\acro{eNB}{eNodeB}
\acro{CN}{Core Network}
\acro{GPRS}{General Packet Radio Service}
\acro{MSC}{Mobile Switching Center}
\acro{BTS}{Base Transceiver Station}
\acro{IMS}{IP Multimedia Subsystem}
\acro{USRP}{Universal Software Radio Peripheral}
\acro{GPSDO}{GPS disciplined oscillator}
\acro{NSA}{Non-standalone}
\acro{U}{Units}
\acro{UBX}{Universal Baseboard Extension}
\acro{HSS}{Home Subscriber Server}
\acro{MME}{Mobility Management Entity}
\acro{DooD}{Docker-outside-of-Docker}
\acro{IMSI}{International Mobile Subscriber Identity}
\end{acronym}

\end{document}